\documentclass[prb]{revtex4}
\usepackage{graphicx}
\usepackage{CJK}
\usepackage{graphicx}

\begin{document}
	\begin{CJK*}{GBK}{song}
		\title{Chaotification via Higher-order Nonlinear Schr\"{o}dinger Equations for Secured Communication}
		\author{Zhenyu Tang} \email{zhenyutang2011@gmail.com}
		\author{Hui-Ling Zhen}
		
		\begin{abstract}
			Higher-order nonlinear Schr\"{o}dinger (HNLS) equation which can be used
			to describe the propagation of short light pulses in the optical fibers,
			is studied in this paper. Using the phase plane analysis, HNLS equation
			is reduced into the equivalent dynamical system, periodicity of such system
			is obtained with the phase projections and power spectra given. By means
			of the time-delay feedback method, with the original dynamical system
			rewritten, we construct a single-input single-output system, and propose
			a chaotic system based on the chaotification of HNLS. Numerical studies
			have been conducted on such system. Chaotic motions with different time
			delays are displayed. Power spectra of such chaotic motions are calculated.
			Lyapunov exponents are given to corroborate that those motions are indeed
			chaotic.
			
		\end{abstract}
		
		
		\maketitle
		
		
		Theoretical and experimental studies on the solitons have revealed that
		the propagation of optical solitons can be used to carry data at a high
		bit rate~\cite{beijing1}. Especially, a source of ultrashort optical
		pulses, which can be used in the high bit rate and long-distance
		optical communication systems, has attracted people's attention~\cite{beijing2,beijing2jia}.
		Techniques have been proposed, including the measurement of ultrafast
		processes~\cite{beijing3}, optoelectronic terahertz time domain
		spectroscopy~\cite{beijing4} and optoelectronic sampling~\cite{beijing5} .
		
		With the consideration of higher-order effects, including the third-order
		dispersion (TOD), self-steepening (SS) and stimulated Raman scattering (SRS)
		in a dispersion-shift fiber, the following higher-order nonlinear
		Schr\"{o}dinger (HNLS) equation reads as~\cite{source1,source2,done3},
		\begin{eqnarray}\label{fangcheng1}
		&& \hspace{-1.5cm} u_x-\frac{i}{2} u_{tt}-i N |u|^2 u+\frac{1}{6}L_H u_{ttt}+S N^2 (|u|^2 u)_t-\tau_R N^2 u |u|^2_t=0,
		\end{eqnarray}
		where $u$, a complex function of the normalized propagation distance $x$ and normalized
		time $t$, refers to the normalized amplitude of the electromagnetic field, the subscripts
		denote the partial derivatives, $N$ and $L_H$ are respective for the nonlinearity and
		dispersion, $S$ and $\tau_R$ account for, respectively, the SS and SRS,
		\begin{eqnarray}\label{fangchengxishu}
		&& \hspace{-1.5cm}  N^2=\frac{\gamma P_0 T_0^2}{|\beta_2|},\ \ \  L_H=\frac{\beta_3}{T_0 \beta_2},
		\ \ \ S=\frac{1}{\omega_0 T_0},\ \ \ \tau_R=\frac{T_R}{T_0},
		\end{eqnarray}
		$P_0$ is the peak power of the incident pulse, $\omega_0$ is the carrier frequency,
		$T_R$ represents the Raman resonant time constant, $\beta_2$ is the group-velocity
		dispersion parameter which is responsible for the pulse broadening, $\beta_3$ is
		the TOD coefficient, $T_0$ is the half-width of the input pulse, and $\gamma$ is
		the fiber nonlinearity coefficient~\cite{source1,source2}. Eq.~(\ref{fangcheng1})
		describe the soliton-effect pulse compression of ultrashort solitons in optical
		fibers~\cite{source1,source2}. Studies on Eq.~(\ref{fangcheng1}) include the numerical
		calculations via the split-step Fourier method~\cite{source1,source2}, bright
		soliton solutions, dynamics of the bright solitons with the random
		initial perturbations, dark soliton solutions~\cite{done3}, stochastic
		soliton solutions~\cite{done4} and rogue wave solutions.
		
		Chaos, which may occur in a deterministic nonlinear system, is a sustained and
		disorderly-looking long-term evolution that satisfies certain  criteria.
		Chaotic motions have been found in some perturbed equations
		and such dynamical systems as the Lorenz system~\cite{lorenz} and Chua's circuit~\cite{chua}.
		With the increasing demands on chaos in such areas as the secure communication
		and information security, the task of making a non-chaotic system chaotic, called the
		anti-control of chaos, or chaotification, has appeared~\cite{anticontrol1,anticontrol2,anticontrol3}.
		
		Such methods as the time-delay feedback~\cite{chaotification1,chaotification2},
		topological conjugate~\cite{chaotify1} and impulsive control~\cite{chaotify2}
		have been developed to chaotify the dynamical systems. A system with time-delay
		is inherently infinitely dimensional, so it can produce the bifurcation and chaos,
		even a first-order system~\cite{timedelay1,timedelay2}. Particularly, chaotic
		motions have been found in some time-delay equations due to their associated
		differential equations~\cite{chaotification1,delayequation}. Therefore, the
		time-delay feedback method has been thought as a straightforward one to chaotify
		a non-chaotic system.
		
		To our knowledge, little work has been completed on Eq.~(\ref{fangcheng1}) with
		respect to the chaos. Motivated by the applications of Eq.~(\ref{fangcheng1})
		in the optical fibers, its potential applications in the secured optical
		communications will be the focus in this paper. Using the time-delay feedback
		method, we aim to propose a new chaotic system with the time-delay introduced
		into Eq.~(\ref{fangcheng1}), and find whether some chaotic motions can be observed.
		Furthermore, when the time-delay is fixed, we want to investigate the possible
		soliton solutions of such chaotic system, and whether such time-delay perturbation
		can affect the soliton propagation.
		
		Methods have been developed to study the nonlinear evolution equations for their
		integrability and periodicity~\cite{spectral}, such as the variational
		approximation~\cite{variational}, phase-plane analysis~\cite{dynamic1,dynamic2,fu1,fu2}
		and perturbation method based on the inverse scattering transform~\cite{inverse1,inverse2}.
		Among them, the phase-plane analysis will be used next.
		
		Setting $u(x,t)=\psi(\xi)e^{i \vartheta}$ with $\xi=a_1 x-b_1 t$ and $\vartheta=a_2 x-b_2 t$,
		and substituting them into Eq.~(\ref{fangcheng1}), we have
		\begin{eqnarray}\label{xitong1}
		&& \hspace{-1.5cm} \psi_{\xi\xi\xi}+r_1 \psi_{\xi\xi}+r_2 \psi_\xi+r_3 \psi+r_4 \psi^3+r_5 \psi^2 \psi_\xi=0,
		\end{eqnarray}
		with
		\begin{eqnarray}\label{xitongxishu}
		&& \hspace{-1.5cm}  r_1=\frac{3 i L_H b_2-3 i}{L_H b_1},\ \ \
		r_2=-\frac{6 a_1-6 b_1b_2+3 L_H b_1 b_2^2}{L_H b_1^3}, \\
		&& \hspace{-1.5cm}  r_3=-\frac{6 i a_2-3 i b_2^2-i b_2^3 L_H}{L_H b_1^3},\ \ \
		r_4=\frac{6 i N^2+6 i b_2 S N^2}{L_H b_1^3},  \\
		&& \hspace{-1.5cm}  r_5=\frac{18 b_1 S N^2-12 \tau_R N^2 b_1}{L_H b_1^3},
		\end{eqnarray}
		where $\psi$ is a real function, $a_j$'s and $b_j$'s ($j=1,2$) are all the real
		constants. Thus, Eq.~(\ref{xitong1}) can be rewritten as a three-dimensional planar
		dynamic system ($X \equiv \psi,Y \equiv \psi_\xi, Z \equiv \psi_{\xi\xi}$),
		\begin{equation}\label{xitong}
		\hspace{-1.7cm} \left\{
		\begin{array}{l}
		\vspace{1.5mm} X_\xi=Y, \\
		\vspace{1.5mm} Y_\xi=Z, \\
		\vspace{1.5mm} Z_\xi=-r_1 Z-r_2 Y-r_3 X-r_4 X^3-r_5 X^2 Y .
		\end{array}
		\right.
		\end{equation}
		
		Phase projections for System~(\ref{xitong}) when $r_2=0$ and $r_2 \neq 0$ are
		shown in Figs.~1(a) and~1(b), respectively. Power spectra for the solutions of
		System~(\ref{xitong}) in the two cases are displayed in Figs.~2(a) and~2(b).
		
		\begin{figure}
			\includegraphics[width=3in]{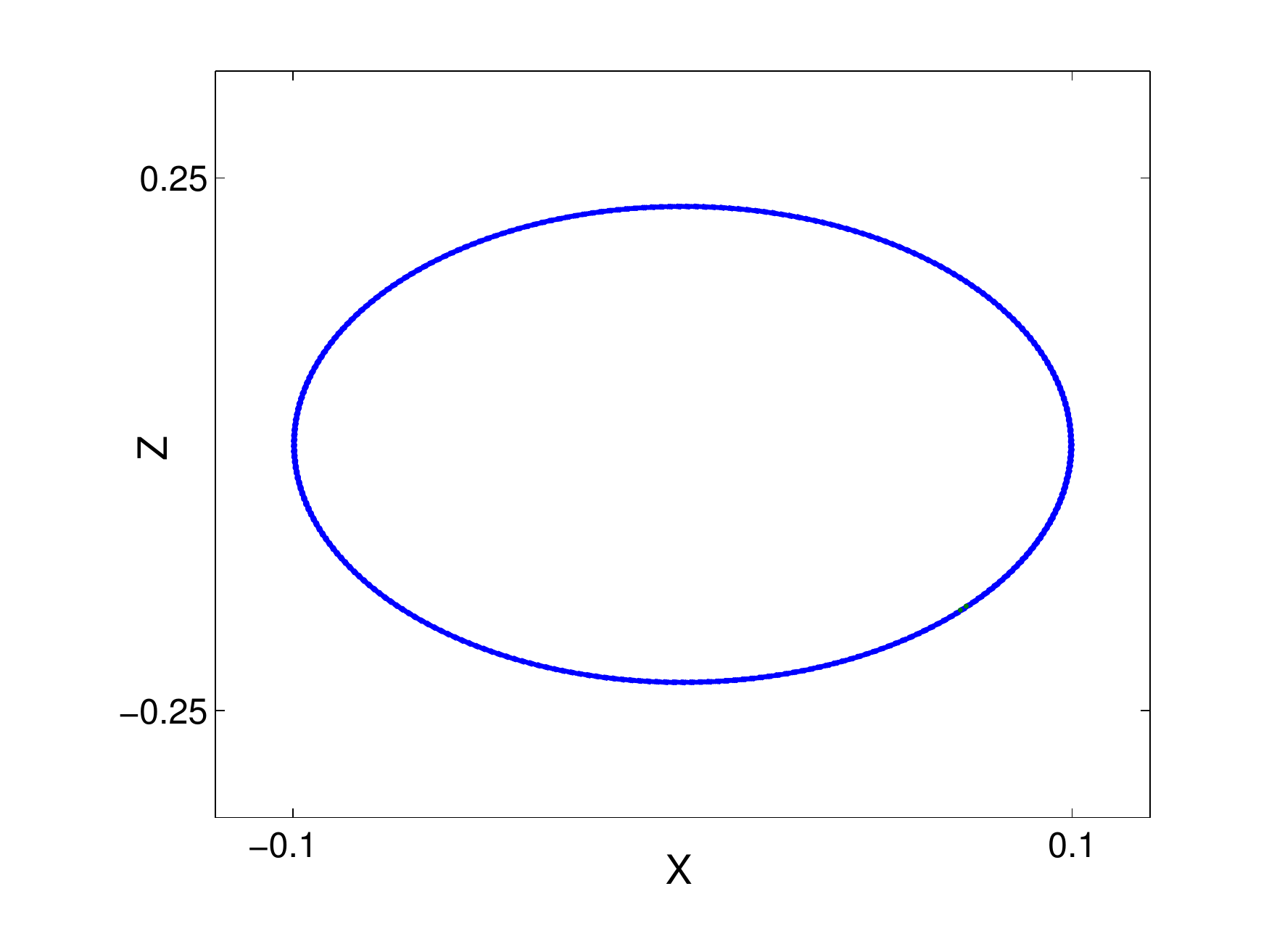}
			\caption{\label{pi}(a) Phase projection of System~(\ref{xitong}) with $r_1$=1, $r_3$=5, $r_4$=2 and $r_5$=1, $r_2$=0.}
		\end{figure}
		
		\begin{figure}
			\includegraphics[width=3in]{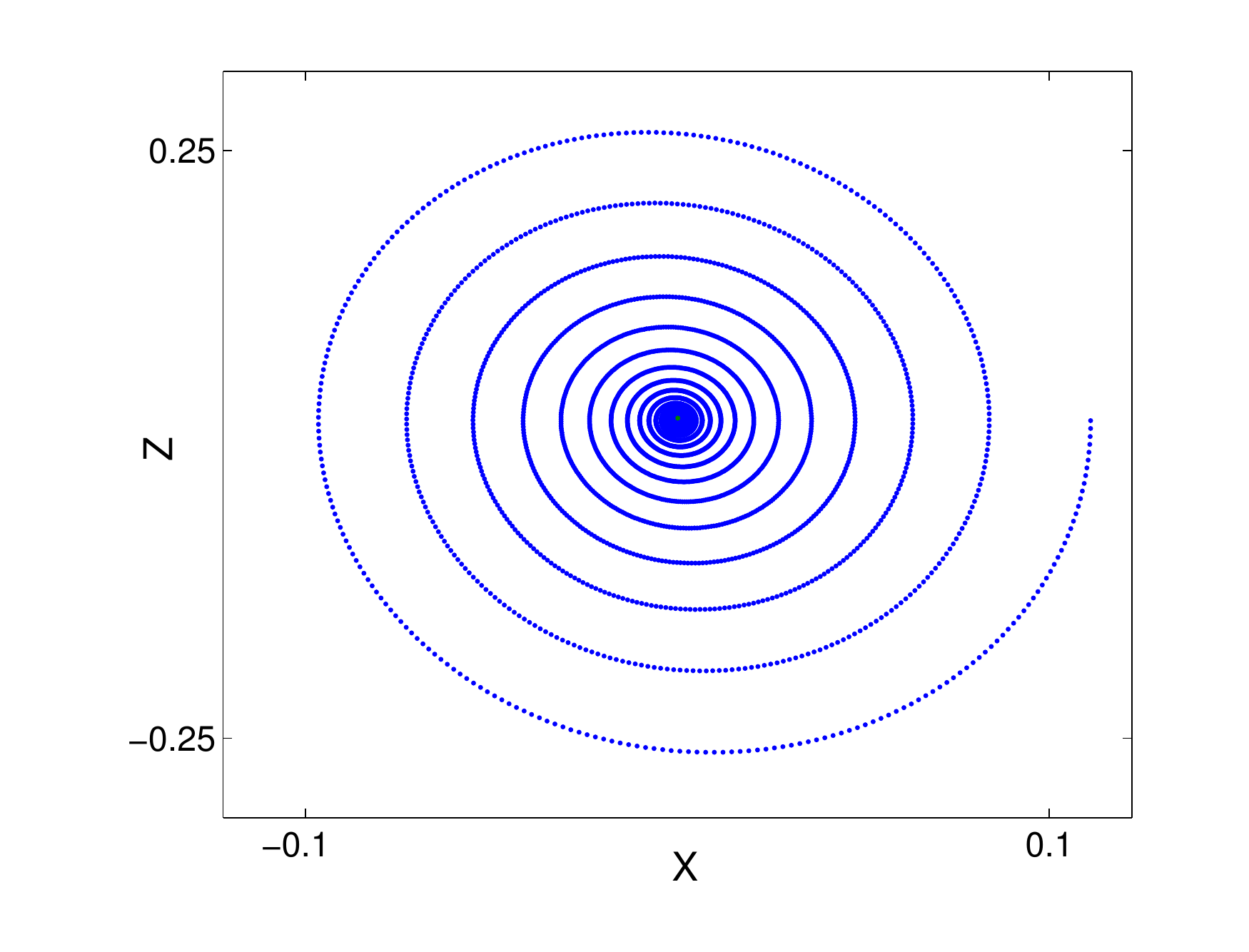}
			\caption{\label{pi}(b) The same as 1(a) but $r_2$=0.2.}
		\end{figure}
		
		\begin{figure}
			\includegraphics[width=3in]{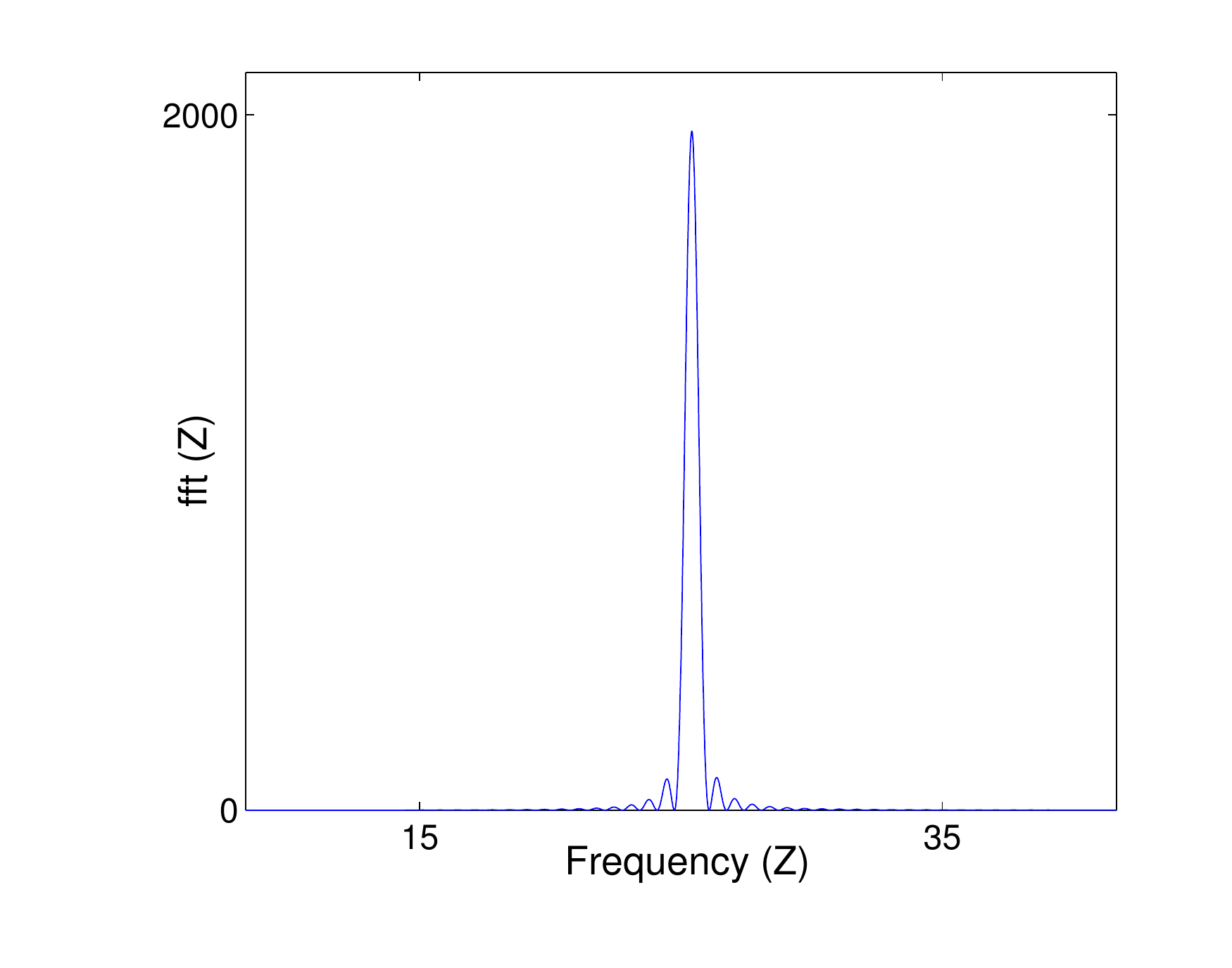}
			\caption{\label{pi}(a) Power spectrum for the solutions of System~(\ref{xitong}) which correspond with Fig.~1(a).}
		\end{figure}
		
		\begin{figure}
			\includegraphics[width=3in]{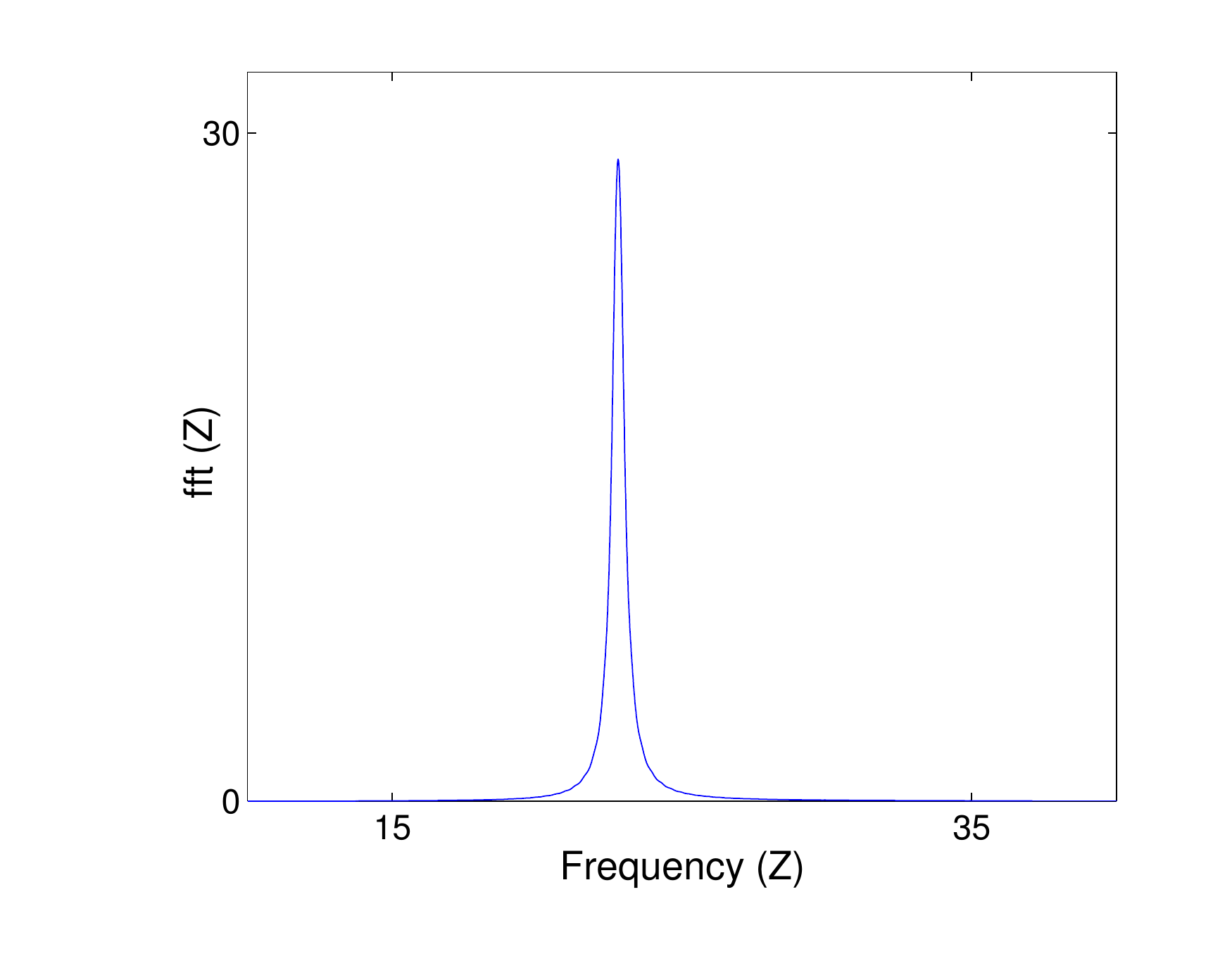}
			\caption{\label{pi}(b) Power spectrum for the solutions of System~(\ref{xitong}) which correspond with Fig.~2(a).}
		\end{figure}
		
		When $r_2=0$, i.e., with the group-velocity dispersion parameter $\beta_2$,
		TOD coefficient $\beta_3$ and half-width of the input pulse $T_0$ satisfying
		that $\beta_3=\frac{(2 b_1 b_2-2 a_1) T_0 \beta_2}{b_1 b_2^2}$, we can see
		a closed curve, which means that there exists a center point, as shown in
		Fig.~1(a). But a focal point can be found in System~(\ref{xitong}) when
		$r_2 \neq 0$, as displayed in Fig.~1(b). Their respective power spectra for the solutions
		of System~(\ref{xitong}) are calculated via the spectral analysis in Figs.~2(a)
		and~2(b), and the periodicity of $Z$ is verified because of the single
		frequency. Hereby, in Figs.~2, the abscissa represents the frequency of
		$Z$ under the certain conditions, where the frequency is the number of
		occurrences of a repeating event per unit time and it is the reciprocal
		of the periodicity~\cite{dynamic2}, and the ordinate $fft(Z)$ represents
		the fast Fourier transform (FFT) of $Z$~\cite{fft}.
		
		System~(\ref{xitong}) can be rewritten as
		\begin{eqnarray}\label{raodongqian}
		\left(\begin{array}{c}
		X \\
		Y \\
		Z
		\end{array} \right)_\xi =\left(\begin{array}{c}
		Y \\
		Z \\
		-r_1 Z-r_2 Y-r_3 X-r_4 X^3-r_5 X^2 Y
		\end{array} \right),
		\end{eqnarray}
		with $\mathbf{x_1}=(0,0,0)^{T}$,
		$\mathbf{x_2}=\left( \sqrt{-\frac{r_3}{r_4}},0,0 \right)^{T}$  and
		$\mathbf{x_3}=\left( -\sqrt{-\frac{r_3}{r_4}},0,0 \right)^{T}$ being its equilibrium
		points, $T$ means the vector transpose. Without loss of generality, we choose $r_2$,
		the coefficient of damped term, as the parameter which can be affected by the
		perturbation via the time-delay feedback method.
		
		To chaotify Eq.~(\ref{fangcheng1}), according to the time-delay feedback
		method~\cite{chaotification1,chaotification2},
		we construct a single-input single-output nonlinear system as follows:
		\begin{eqnarray}\label{raodong1}
		&& \hspace{-1.5cm}  \mathbf{x}_\xi=\mathbf{f}(\mathbf{x})+\mathbf{g}(\mathbf{x})\delta(\xi),\label{siso1a}\\
		&& \hspace{-1.5cm}  y=h(\mathbf{x}),
		\end{eqnarray}
		where $\mathbf{x}$ and $y$ label the input and output, respectively,
		$\textbf{x}_\xi=d \textbf{x}/d \xi$, $\mathbf{f}(\mathbf{x})$ and $\mathbf{g}(\mathbf{x})$
		are both real vector functions, $\delta(\xi)$ corresponds to a system parameter perturbation,
		and $h(\mathbf{x})$ is a smooth real function and refers
		to the output. In the case of System~(\ref{raodongqian}), $\mathbf{x}=(X,Y,Z)^{'}$,
		$\mathbf{f}(\mathbf{x})$ and $\mathbf{g}(\mathbf{x})$ can be expressed as
		\begin{eqnarray}\label{fg}
		&& \hspace{-1.5cm} \mathbf{f}(\mathbf{x})=\left(\begin{array}{c}
		Y \\
		Z \\
		-r_1 Z-r_2 Y-r_3 X-r_4 X^3-r_5 X^2 Y
		\end{array} \right),\ \ \
		\mathbf{g}(\mathbf{x})=\left(\begin{array}{c}
		0 \\
		0 \\
		Y
		\end{array} \right).
		\end{eqnarray}
		Thus, System~(\ref{raodong1}) can be rewritten as
		\begin{eqnarray}\label{raodong2}
		&& \hspace{-1.5cm}  \mathbf{x}_\xi=\left(\begin{array}{c}
		Y \\
		Z \\
		-r_1 Z-r_2 Y-r_3 X-r_4 X^3-r_5 X^2 Y
		\end{array} \right)+\left(\begin{array}{c}
		0 \\
		0 \\
		Y
		\end{array} \right) \delta(\xi),\label{siso1a}\\
		&& \hspace{-1.5cm}  y=h(\mathbf{x}),
		\end{eqnarray}
		where $h(\mathbf{x})$ can be determined based on $\delta(\xi)$.
		
		Based on Expressions~(\ref{fg}), we have
		\begin{eqnarray}\label{ad}
		&& \hspace{-1.5cm} \mathbf{ad_f g}(\mathbf{x})=\left(\begin{array}{c}
		0 \\
		-Y \\
		r_1 Y+Z
		\end{array} \right),\ \ \
		\mathbf{ad_f}^2 \mathbf{g}(\mathbf{x})=\left(\begin{array}{c}
		Y \\
		-r_1 Y-Z \\
		(r_1^2-r_2) Y-r_3 X-r_4 X^3-r_5 X^2 Y
		\end{array} \right),\nonumber
		\end{eqnarray}
		where $\mathbf{ad_f g}(\mathbf{x})$ is the Lie bracket of the two smooth vector functions
		$\mathbf{f}(\mathbf{x})$ and $\mathbf{g}(\mathbf{x})$.
		Based on the conclusions in
		Refs.~\cite{chaotification1,chaotification2},
		the relative degree of System~(\ref{raodong1}) is three, i.e., the dimension
		of $\mathbf{x}$. Hereby, the definition of ``relative degree" can be seen in
		Refs.~\cite{chaotification1,chaotification2},
		and ``Lie bracket" $\mathbf{ad_f g}(\mathbf{x})$ and
		$\mathbf{ad_f}^2 \mathbf{g}(\mathbf{x})$ can be obtained
		as~\cite{chaotification1,chaotification2}.
		\begin{eqnarray}
		&& \hspace{-1cm} \mathbf{ad_f g}(\mathbf{x})=\frac{\partial \mathbf{f}(\mathbf{x})}{\partial \mathbf{x}} \mathbf{g}(\mathbf{x}) -\mathbf{f}(\mathbf{x}) \frac{\partial \mathbf{g}(\mathbf{x})}{\partial \mathbf{x}},\nonumber \\
		&& \hspace{-1cm} \mathbf{ad_f}^2 \mathbf{g}(\mathbf{x})=\mathbf{ad_f}[\mathbf{ad_f g}(\mathbf{x})]
		=\frac{\partial \mathbf{f}(\mathbf{x})}{\partial \mathbf{x}} [\mathbf{ad_f g}(\mathbf{x})] -\mathbf{f}(\mathbf{x}) \frac{\partial [\mathbf{ad_f g}(\mathbf{x})]}{\partial \mathbf{x}}.\nonumber
		\end{eqnarray}
		
		Via the time-delay feedback method~\cite{chaotification1,chaotification2},
		we know that $h(\mathbf{x})$ should satisfy
		\begin{eqnarray}
		&& \hspace{-1.5cm} \frac{\partial h(\mathbf{x})}{\partial \mathbf{x}}
		\left[\mathbf{g}(\mathbf{x}),\mathbf{ad_f g}(\mathbf{x}),\mathbf{ad_f}^2 \mathbf{g}(\mathbf{x})\right]=0, \nonumber
		\end{eqnarray}
		i.e.,
		\begin{eqnarray}\label{jisuan2}
		&& \hspace{-1.5cm} \frac{\partial h(\mathbf{x})}{\partial \mathbf{x}} \mathbf{g}(\mathbf{x})
		=\frac{\partial h}{\partial Z} Y=0,  \\
		&& \hspace{-1.5cm} \frac{\partial h(\mathbf{x})}{\partial \mathbf{x}} \mathbf{ad_f g}(\mathbf{x})
		= \frac{\partial h}{\partial Y} (-Y)+\frac{\partial h}{\partial Z} (r_1 Y+Z)=0,
		\end{eqnarray}
		so that we have $h(\mathbf{x})=X$. In line with
		Refs.~\cite{chaotification1,chaotification2},
		$\delta(\xi)$ can be expressed as
		\begin{eqnarray}
		&& \hspace{-1.5cm} \delta(\xi)=\varsigma \sin[\sigma X(\xi-\tau)],
		\end{eqnarray}
		with $\varsigma$ and $\sigma$ being both the real constants,
		$\tau$ refering to the time delay, and $\xi$ given in Sec.~2.
		
		As a generalization of this part, the chaotification of
		Eq.~(\ref{fangcheng1}) can be given as
		\begin{eqnarray}\label{raodonghou}
		\mathbf{x}_\xi= \left(\begin{array}{c}
		X \\
		Y \\
		Z_1
		\end{array} \right)_\xi =\left(\begin{array}{c}
		Y \\
		Z \\
		-r_1 Z-(r_2+\delta) Y-r_3 X-r_4 X^3-r_5 X^2 Y
		\end{array} \right),
		\end{eqnarray}
		with $\delta=\delta(\xi)=\varsigma \sin[\sigma X(\xi-\tau)]$
		being the time-delay perturbation on $r_2$.
		
		To study the possible chaotic motions of System~(\ref{raodonghou}),
		we will investigate the phase projections of $Z$ in Figs.~3. Note
		that the difference between Figs.~3(a) and~3(b) roots in the
		different values of the time delay $\tau$. Then, we calculate
		their respective power spectra in Figs.~4, which can be used for people
		to corroborate that the motions in Figs.~3 are indeed chaotic.
		
		\begin{figure}
			\includegraphics[width=3in]{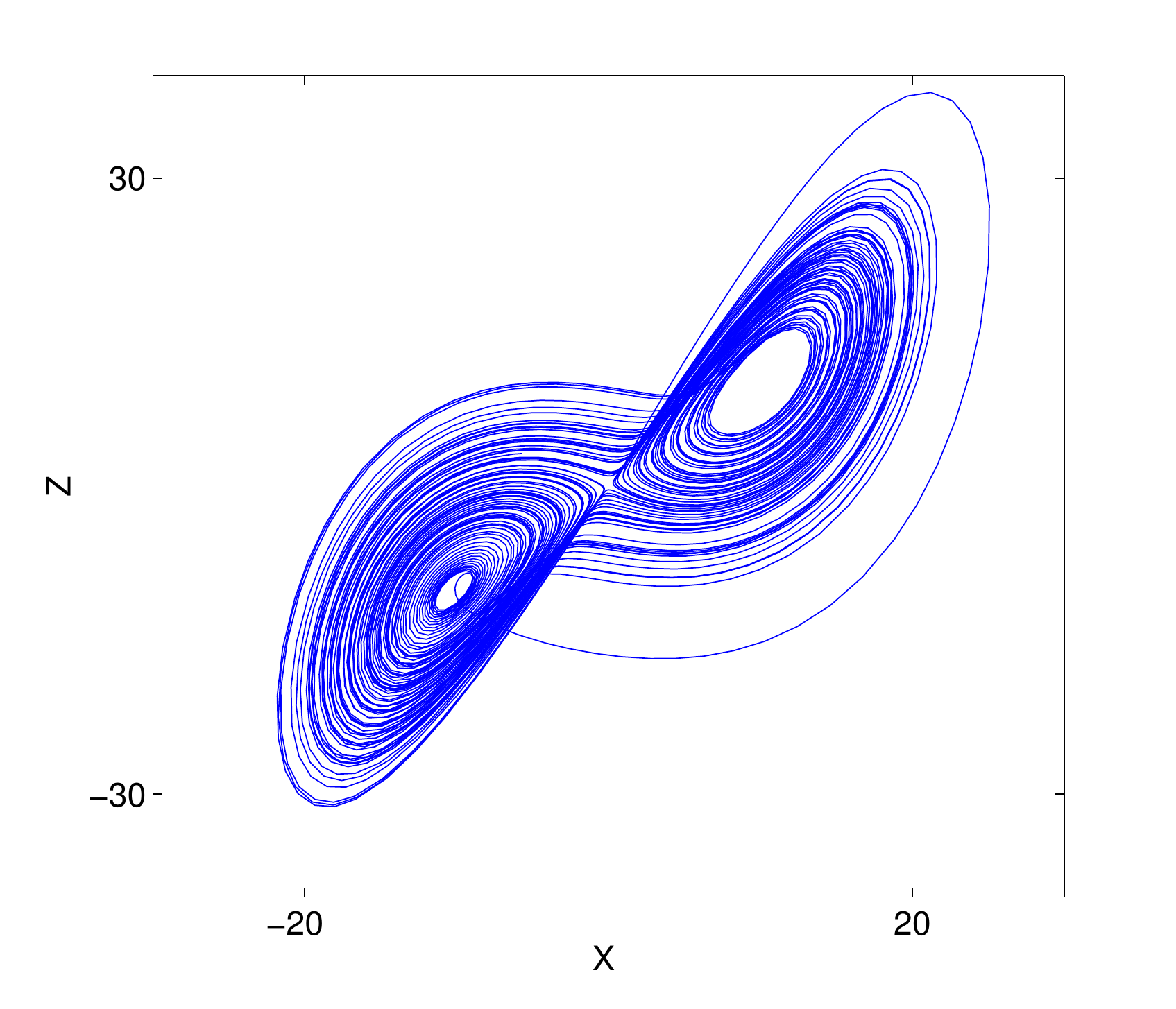}
			\caption{\label{pi}(a) Phase projection of System~(\ref{raodonghou}) with $r_1$=1.5,
				$r_2$=5, $r_3$=-1, $r_4$=5, $r_5$=1.75,
				$\varsigma$=1, $\sigma$=0.5, $\tau$=10.}
		\end{figure}
		
		\begin{figure}
			\includegraphics[width=3in]{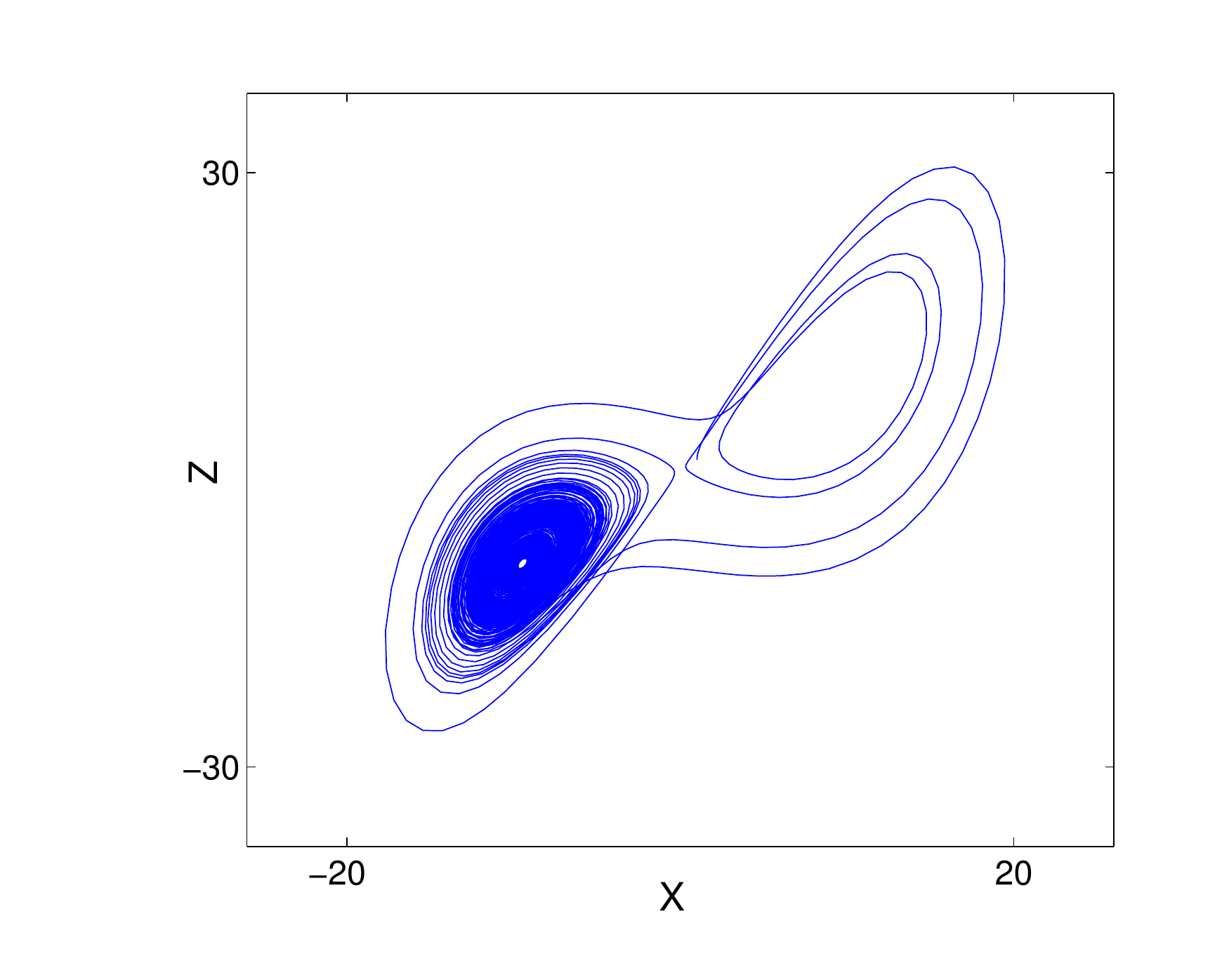}
			\caption{\label{pi}(b) Phase projection of System~(\ref{raodonghou}) with $r_1$=1.5,
				$r_2$=5, $r_3$=-1, $r_4$=5, $r_5$=1.75,
				$\varsigma$=1, $\sigma$=0.5, $\tau$=26.}
		\end{figure}
		
		\begin{figure}
			\includegraphics[width=3in]{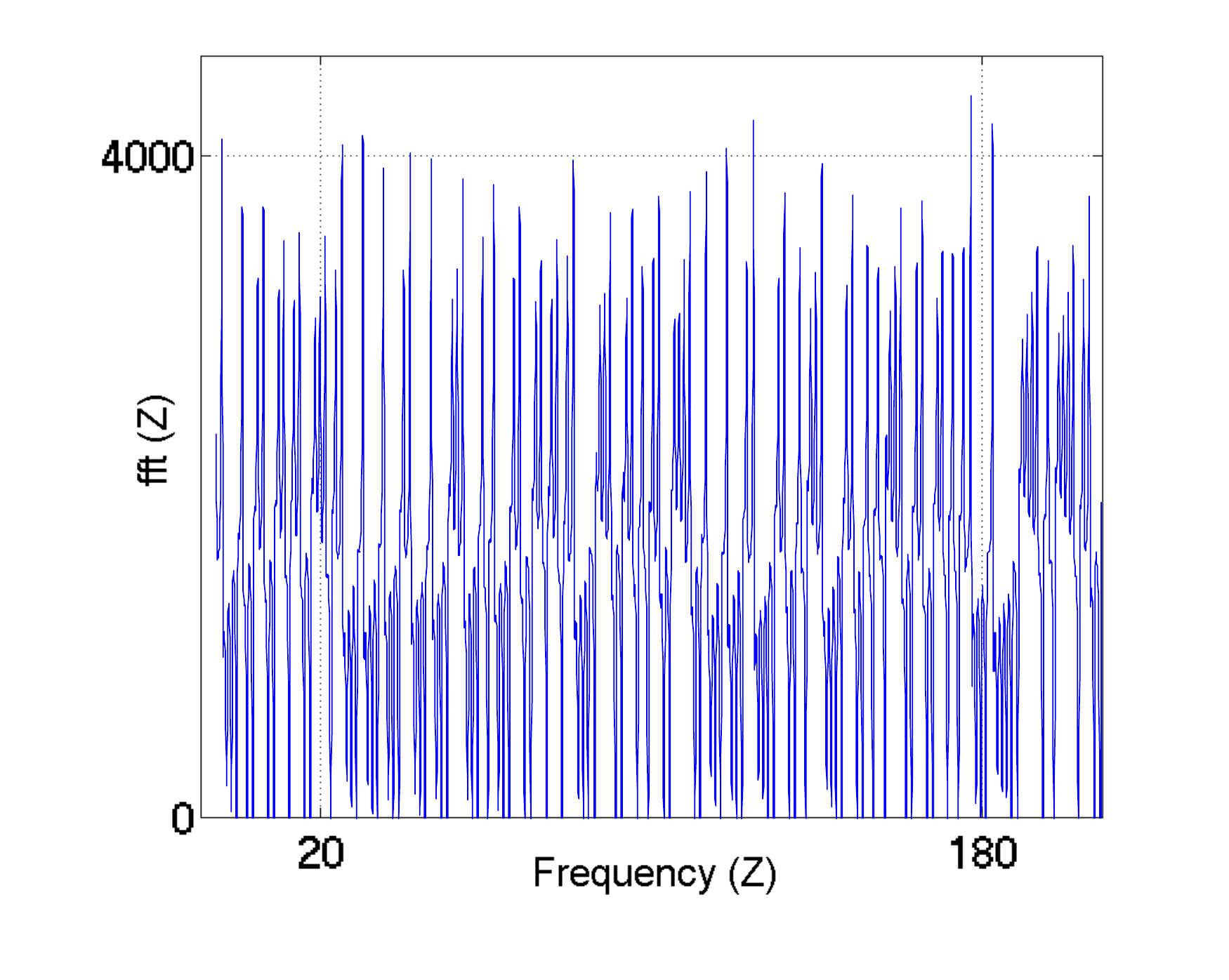}
			\caption{\label{pi}(a) Power spectrum for the solutions of System~(\ref{raodonghou})
				which correspond with Fig.~3(a).}
		\end{figure}
		
		\begin{figure}
			\includegraphics[width=3in]{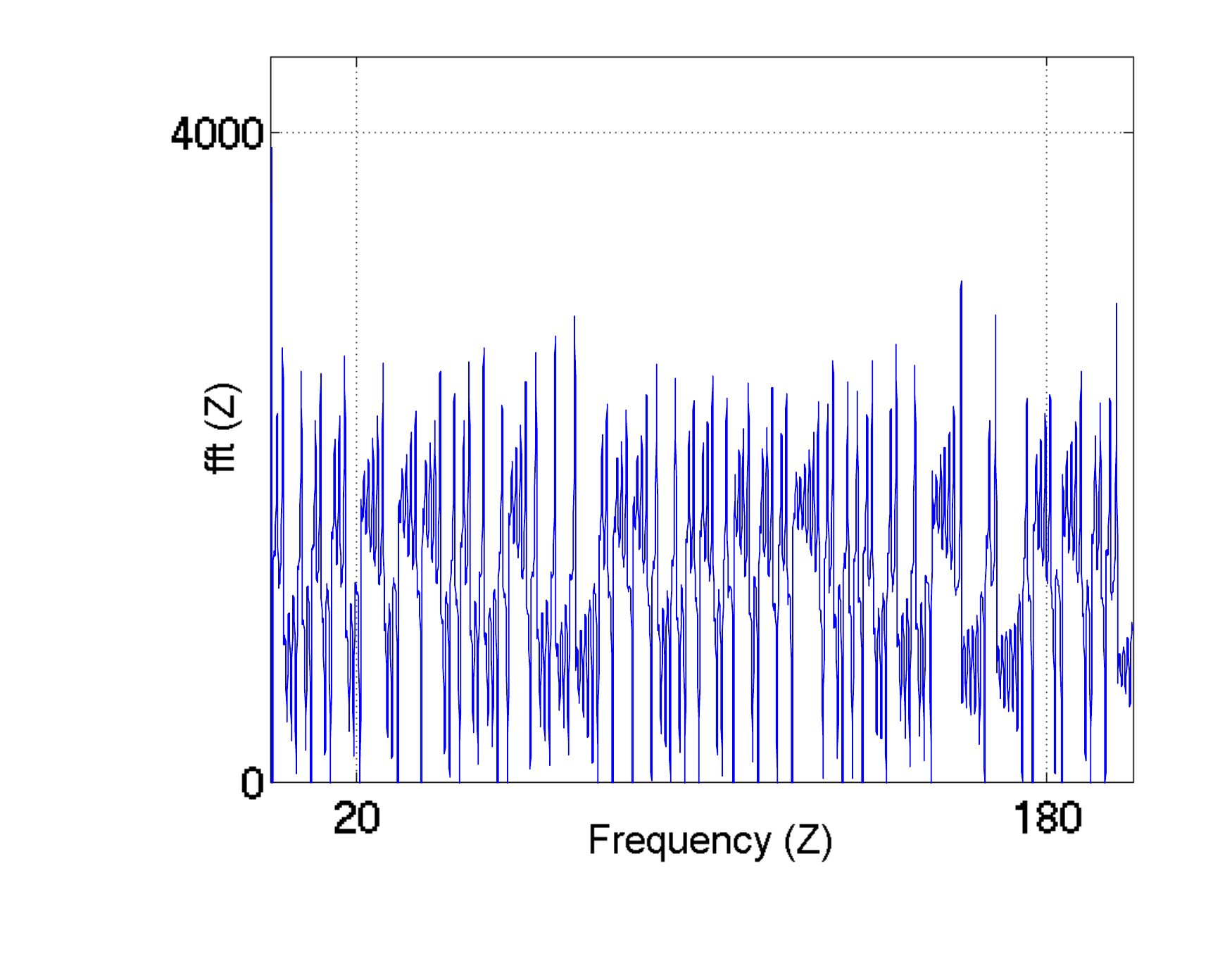}
			\caption{\label{pi}(b) Power spectrum for the solutions of System~(\ref{raodonghou})
				which correspond with Fig.~4(a).}
		\end{figure}
		
		Comparing Figs.~3-4 with Figs.~1-2, we can see that there exits no
		periodicity for System~(\ref{raodonghou}), and the frequency of
		System~(\ref{xitong}) has been broken. As defined in Refs.~\cite{developed1,developed2},
		developed chaos occurs when the solutions ignore the driver periods
		and represent a random sequence of uncorrelated shocks. Thus,
		developed chaos occurs in Figs.~3.
		
		To verify that the behaviors in Figs.~3 are indeed chaotic
		from another point of view, we will investigate the Lyapunov
		exponents~\cite{theory1,theory2} of the time delay $\tau$
		and parameter $r_2$, as displayed in Figs.~5.
		
		\begin{figure}
			\includegraphics[width=3in]{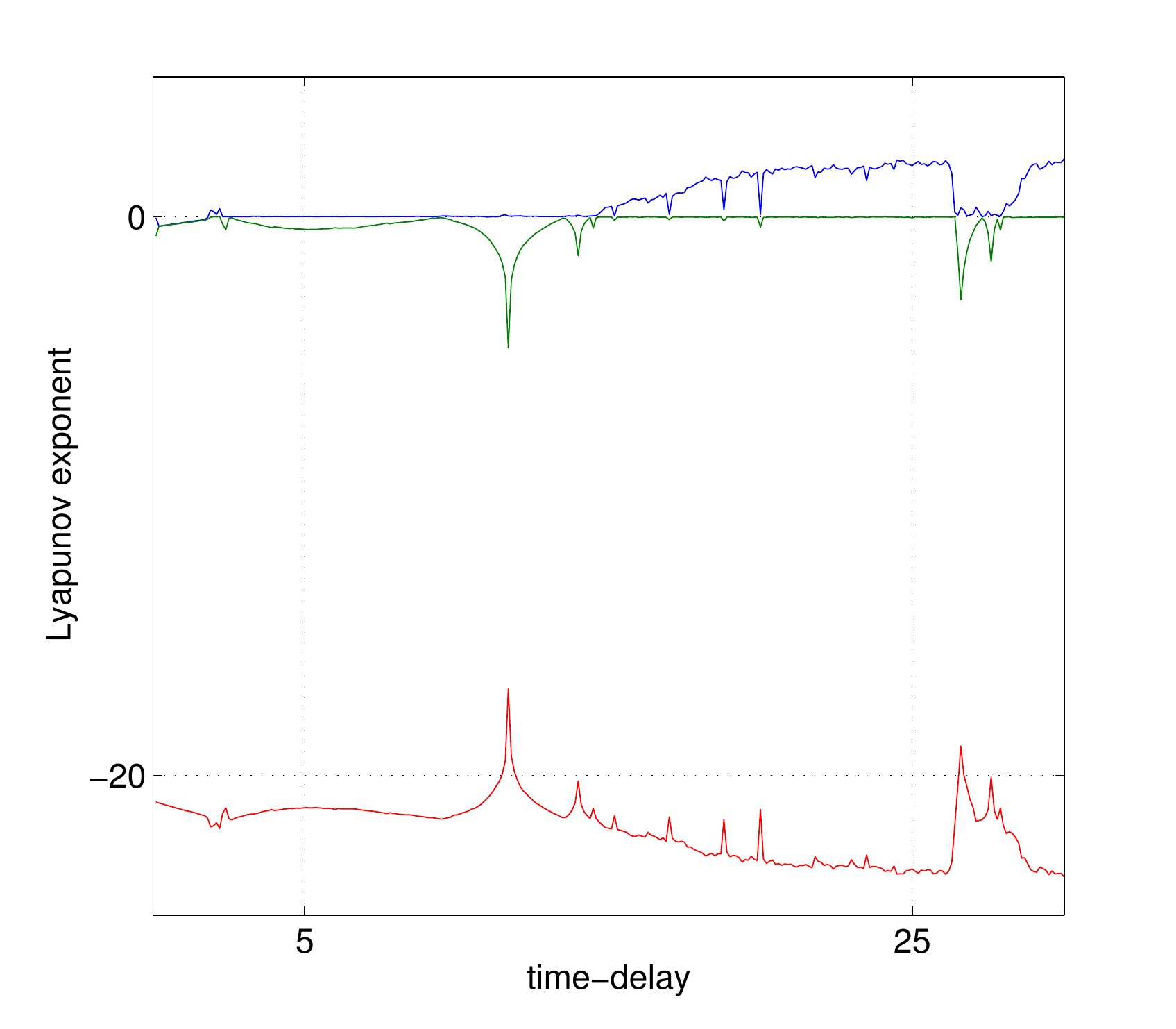}
			\caption{\label{pi}(a) Lyapunov exponents when the time-delay $\tau$ is the bifurcation parameter, while
				$r_1$=1.5, $r_2$=2, $r_3$=-1, $r_4$=5, $r_5$=1.75, $\varsigma$=1 and $\sigma$=0.5.}
		\end{figure}
		
		\begin{figure}
			\includegraphics[width=3in]{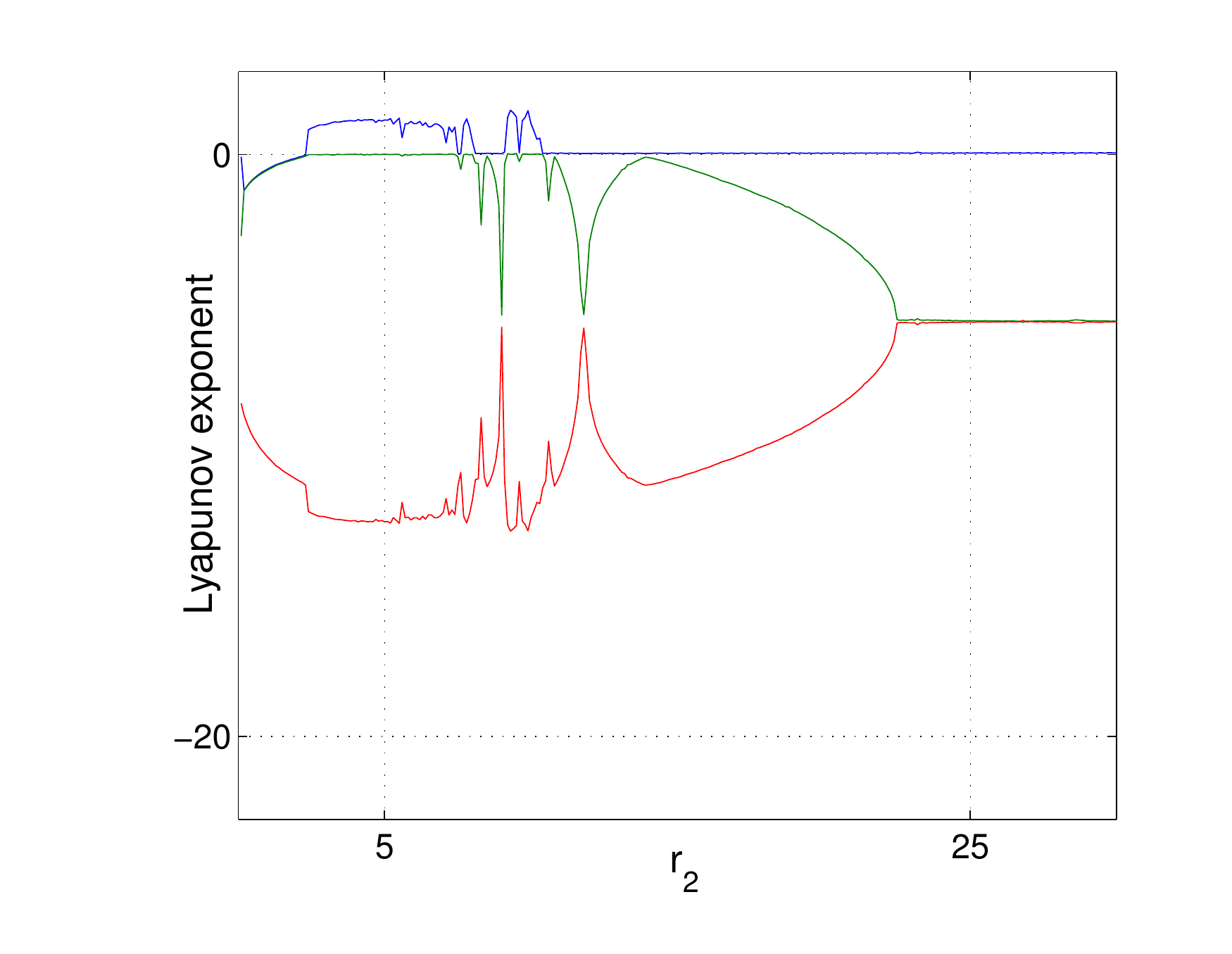}
			\caption{\label{pi}(b) Lyapunov exponents when $r_2$ is the bifurcation parameter, while
				$r_1$=1.5, $\tau$=20, $r_3$=-1, $r_4$=5, $r_5$=1.75, $\varsigma$=1 and $\sigma$=0.5.}
		\end{figure}
		
		Dynamic behaviors of System~(\ref{raodonghou}) varying with $\tau$
		can be found in Fig.~5(a). We can see that System~(\ref{raodonghou})
		can give rise to the periodic motions when $2.5 \leq \tau \leq 15$
		or $26.5 \leq \tau \leq 28$ with other parameters fixed, while
		System~(\ref{raodonghou}) may turn into the chaos when $15 < \tau <26.5$
		or $28 < \tau < 30$. Dynamic behaviors of System~(\ref{raodonghou})
		varying with $r_2$ are shown in Fig.~5(b). Note that System~(\ref{raodonghou})
		turns into the chaos when $2.5 < r_2 <11$, and System~(\ref{raodonghou})
		is periodic when $11 \leq r_2 \leq 30$. As expected, the Lyapunov exponents
		in Figs.~5 are consistent with the chaotic motions shown in Figs.~3.
		It also shows that Eq.~(\ref{fangcheng1}) can be chaotified via
		the time-delay method when the parameters are in certain ranges.
		
		In this paper, we have focused ourselves on a HNLS equation [i.e., Eq.~(\ref{fangcheng1})]
		in the secured optical communication, which can be used to describe
		the propagation of short light pulses in the optical fibers.
		With the time-delay perturbation introduced, we have proposed a chaotic
		system [i.e., System~(\ref{raodonghou})], which has been studied
		numerically with the phase projections, power spectra and Lyapunov exponents.
		Furthermore, to study the potential applications of Eq.~(\ref{fangcheng1})
		in the secured optical communication from another point of view,
		we have investigated the soliton solutions of System~(\ref{raodonghou})
		when the time delay is fixed. The main results can be summarized
		as below:
		
		$\bullet$ Using the phase plane analysis, we have given the periodicity
		of Eq.~(\ref{fangcheng1}), as seen in
		Figs.~1 and~2. Effect of damped coefficient $r_2$ has been obtained,
		i.e., center point has been shown in Fig.~1(a) if $r_2=0$, or focal
		point in Fig.~2(a).
		
		$\bullet$ Making use of the time-delay feedback method, we have
		chaotified Eq.~(\ref{fangcheng1}), and a chaotic system [i.e., System~(\ref{raodonghou})]
		has been constructed with the time-delay perturbation introduced into Eq.~(\ref{fangcheng1}).
		
		$\bullet$ Phase projections of System~(\ref{raodonghou}) with different
		time delays have been given in Figs.~3(a) and~3(b), and their respective
		power spectra have been displayed in Figs.~4(a) and~4(b).
		
		$\bullet$ Lyapunov exponents of the time-delay $\tau$ and parameter $r_2$
		have been shown in Figs.~5(a) and~5(b), respectively. We have found
		that when $r_2$ is fixed, System~(\ref{raodonghou}) is periodic when
		$2.5 \leq \tau \leq 15$ or $26.5 \leq \tau \leq 28$, while it may turn
		into the chaos when $15 < \tau <26.5$ or $28 < \tau < 30$. When $\tau$
		is fixed, System~(\ref{raodonghou}) is chaotic when $2.5 < r_2 <11$, and it
		is periodic when $11 \leq r_2 \leq 30$.
		
		\noindent\textbf {Acknowledgments}
		The authors acknowledge *** for the discussions during the works.

	\end{CJK*}
\end{document}